\documentclass[twocolumn,preprintnumbers,elsart]{revtex4}
\usepackage{amsmath}

\usepackage{mathrsfs}
\usepackage{graphicx}
\usepackage{dcolumn}
\usepackage{bm}
\usepackage[center]{subfigure}

\begin{document}

\title{Slow-light all-optical soliton diode based on tailored Bragg-grating structure }

\author {Hongji Li}
\affiliation{Department of Applied Physics, South China
Agricultural University, Guangzhou 510642, China}
\author{Zhigui Deng}
\affiliation{Department of Applied Physics, South China
Agricultural University, Guangzhou 510642, China}
\author{Jiasheng Huang}
\affiliation{Department of Applied Physics, South China
Agricultural University, Guangzhou 510642, China}
\author{Shenhe Fu}
\affiliation{State Key Laboratory of Optoelectronic Materials and Technologies,
Sun Yat-sen University, Guangzhou 510275, China.}
\author{Yongyao Li}\email{Corresponding author: yongyaoli@gmail.com}
\affiliation{Department of Applied Physics, South China
Agricultural University, Guangzhou 510642, China}

%
%

\begin{abstract}An all optical soliton diode (AOSD) is proposed based on a sandwich nonlinear Bragg-grating structure: a linearly chirped Bragg-grating linked to a uniform Bragg-grating and again to a chirped Bragg-grating.  The nonreciprocity is achieved by introducing two spatially asymmetric chirped Bragg-gratings with optical nonlinearity.  High transmission ratio up to 150 is obtained when launching a picosecond Gaussian pulse into the setting. We find that such pulses in a form of solitons propagate at a rather small velocity ($<0.03c$) when pulse wavelength is selected in the vicinity of photonic bandgap. 
\end{abstract}


\maketitle 

Optical diode, a nonreciprocal device that transmits light only in one direction and blocks that in the opposite direction, is a key enabler for optical information processing. To realize the optical diode, breaking the symmetry of wave propagation is necessary but it is still a challenge problem. Many schemes have been proposed to solve such a problem. Conventional methods were based on the Faraday effect in magneto-optical crystals \cite{Carson1964}, but such magneto-optical component hinders the applications in integrated photonic systems. Since then, the research on optical diodes became very intense, and many compact schemes were proposed, using opto-acoustic effect \cite{Kang2011}, second harmonic generation \cite{Carlos1996}, absorbing multilayer system \cite{Gevorgyan2002}, asymmetric nonlinear absorption \cite{Philip2007}, ring-resonator structures \cite{Fan2009,Fan2012}, Fano effect \cite{Wding2012,Xuyi2014}, micro-optomechanical devices with flexible Fabry-perot \cite{Lipson2009}, left-hand materials \cite{Feise2005} and photonic crystal structures \cite{CXue2010,Liu2012,Wang2013,Zhang2014,Bulgakow2014}. While majorities of these work concentrated on high nonreciprocal transmission ratio \cite{Fan2012,Zhang2014}, the shape of the optical field after propagating through the devices was neglected.

In addition to those attempts, an important candidate for the generation of optical diodes is based on periodic grating structures due to their tunability and cascadability. Using the edge of photonic bandgap \cite{Bloemer1994}, as well as the consideration of optical nonlinearity, the optical diode was first put forward in a one-dimensional photonic material with spatial gradation in the refractive index. Later, an all-optical diode was demonstrated in a periodically LiNbO$_{3}$ channel waveguide using quasi-phase matching \cite{Assanto2001}. Non-reciprocal transmission of circular polarized light at the photonic-bandgap regions based on electro-tunable liquid-crystal hetero-junctions was demonstrated experimentally and theoretically \cite{Hwang2005}.

Similarly to \cite{Hwang2005}, we propose a sandwich structure based on nonlinear Bragg gratings: a linearly chirped Bragg-grating (CBG) \cite{CBG} linked through a uniform Bragg-graitng (UBG), and again to a CBG. Using this setting, a hight-Q and slow-light all-optical soliton diode (AOSD) is achieved. It was demonstrated that, by introducing a CBG cascaded a UBG structure \cite{Fu1,Fu2}, strong retardation of optical pulses, as well as the creation of standing stable pulses were realized. These pulses can preserve its shape while propagating along the gratings, as a results of the balance between optical induced nonlinearity and pulse dispersion, which can be termed as a BG soliton \cite{BG1,BG2,BG3,BG33,BG4,BGT,BG5,BG6,BG7}. In our setting, it is possible to break the reciprocity by introducing such two spatially asymmetric CBG structures, linked via a UBG segment. The paper is organized as following, we firstly present the standard nonlinear coupled-mode theory to discuss the possibility for the realization of the nonreciprocity of the device. Then we apply such nonreciprocity to realize an AOSD. Performance of such diodes is systematically discussed. Conclusions are made in the end of this paper.

\begin{figure}[tbp]
\centering{\label{fig1a}
\includegraphics[scale=0.25]{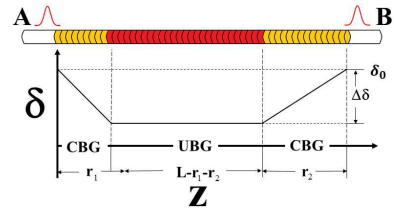}}%
\caption{(Color online) An AOSD sketch constructed by a sandwich structure: CBG+UBG+CBG.}
\label{Model}
\end{figure}

Fig. \ref{Model} shows the sketch of the sandwich structure. The total length of the setting is $L$, and the lengths of two CBG segments are $r_{1}$ and $r_{2}$, respectively. The linear refractive index \cite{Fu1,Fu2}: $n(z)=n_{0}[1+2\Delta n\cos(2\pi z(1+C_{1}z)/\Lambda_{0})]$ for the left CBG segment (i.e., $z\in[0,r_{1}]$); $n(z)=n_{0}[1+2\Delta n\cos(2\pi z(1+C_{1}r_{1}/\Lambda_{0})]$ for the UBG segment (i.e., $z\in(r_{1},L-r_{2})$); $n(z)=n_{0}[1+2\Delta n\cos(2\pi z[1+C_{2}(L-z)]/\Lambda_{0})]$ for the right CBG segment (i.e., $z\in[L-r_{2},L]$). Here, $z$ is the propagation distance. $C_{1}$ and $C_{2}$ are the chirp coefficients of these two CBG segments, respectively. $\Lambda_{0}$ is the BG period at the input and output edges. $n_{0}$ is the average refractive index, and $\Delta n$ is the amplitude of the refractive-index modulation.

The BG solitons propagating through the setting are well described by the following nonlinear coupled-mode equations (CMEs) \cite{Fu1,Fu2,SongLY}
\begin{eqnarray}
&&\pm i{\partial E_{f,b} \over\partial z}+{i\over v_g}
{\partial E_{f,b} \over\partial t}+
\delta(z)E_{f,b}+\kappa(z)E_{b,f}\nonumber\\
&&+\gamma(|E_{f,b}|^{2}+2|E_{b,f}|^{2})E_{f,b}=0, \label{basicEq}
\end{eqnarray}
where $t$ is time, $E_f$ and $E_b$ are amplitudes of the forward-traveling and backward-traveling waves of the soliton. $v_g=c/n_0$
is the linear group velocity in the material of which the BG is fabricated, with $c$ being the light speed in the vacuum, and $\gamma=n_{2}\omega/c$ is the nonlinearity coefficient, with
$\omega$ being the carrier frequency and $n_{2}$ the Kerr coefficient.
The coupling strength between forward and backward waves is $\kappa=\pi\Delta n/\Lambda_{0}$, while $\delta(z)$, which represents the local
wavenumber detuning, is
\begin{eqnarray}
\delta(z)=\begin{cases}
\delta_{0}-2\pi C_{1}z/\Lambda_{0} & z\in[0,r_{1}),\\
\delta_{0}-\Delta\delta & z\in[r_{1},L-r_{2}],\\
\delta_{0}-2\pi C_{2}(L-z)/\Lambda_{0} & z\in(L-r_{2},L],
\end{cases} \label{deltaz}
\end{eqnarray}
where $\delta_{0}\equiv2\pi n_{0}/\lambda - \pi/\Lambda_{0}$ is the detuning at the left and right edges of the CBG segment, and $\Delta\delta\equiv2\pi C_{1}r_{1}/\Lambda_{0}=2\pi C_{2}r_{2}/\Lambda_{0}$, which is a constant, is the detuning difference between the UBG segment and the left/right edge of the setting.

\begin{figure}[tbp]
\centering{\label{fig2a}
\includegraphics[scale=0.3]{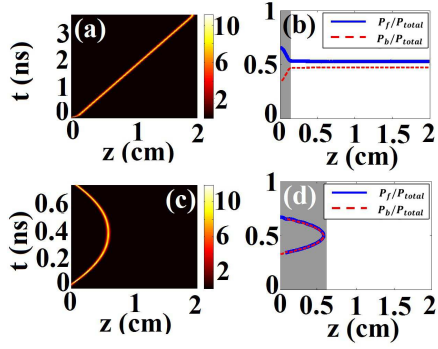}}%
\caption{(Color online) Typical examples of pulse propagation for two different length of CBG, $r_{1}=0.1$ cm (a) and $r_{1}=0.6$ cm (c), with the same PI, $I=4.24$ GW/cm$^{2}$. (b,d)$P_{\mathrm{f}}$ and $P_{\mathrm{b}}$ versus $z$, corresponding to (a,c) respectively. In (b,d), the gray areas are the CBG regions and the blue solid curves are the $P_{\mathrm{f}}(z)/P_{\mathrm{total}}$, while the red dash curves are  $P_{\mathrm{b}}(z)/P_{\mathrm{total}}$. Here $P_{\mathrm{total}}=\int^{\infty}_{-\infty}(|E_{f}|^{2}+|E_{b}|^{2})dt$. and $L$ is fixed to 2 cm.}
\label{SM1}
\end{figure}

Simulations of Eq. (\ref{basicEq}) were performed by means of the fourth-order Runge-Kutta method, with physical parameters of the silicon, whose self-focusing Kerr coefficient is $4.5\times10^{-14}$ cm$^{2}/$W \cite{TPA}, and the average refractive index is $n_{0}=3.42$. The refractive index modulation depth is taken as $\Delta n=0.006$. To avoid the two-photon absorption \cite{TPA}, the carrier wavelength of $1798.4$ nm is selected, and the BG period at the input edge of the sample is fixed to $\Lambda_{0}=263.16$ nm, giving rise to $\delta_{0}=106.2$ cm$^{-1}$. At the input edges of the sample, a picosecond Gaussian pulse, with a temporal width 3.1 ps and a tunable peak intensity (PI), namely $I$, is coupled into the system.

As mentioned \cite{Fu2}, the introduction of CBG segment decreases the pulse propagating group velocity, thus affecting the optical transmission. To see this clearly, we fix $r_{2}=0$ and launch the pulse from the left edge of the sample (i.e. side A). Moreover, we fix $\Delta\delta\equiv7.31$ cm$^{-1}$, so that the considered wavelength is located at the edge of the photonic bandgap (here, the chirp coefficient, $C_{1}=\Delta\delta\Lambda_{0}/2\pi r_{1}$, is a function of $r_{1}$). Typical examples of the pulse transmission and reflection with two different values of $r_{1}$ are displayed in Fig. \ref{SM1} (a,c), under identical incident intensity. From Fig. \ref{SM1}(a), it is observed that the pulses maintain its shape and pass through the setting. It indicates that strong optical nonlinearity is induced, overcoming the pulse dispersion near the edge of bandgap. As a result, a BG soliton is generated propagating in UBG segment. Moreover, as expected, the pulse cannot penetrate the setting with a larger CBG length, see Fig. \ref{SM1}(c), eventually bounced back to the incident edge. These assertions can be verified by the comparison between these two characters: $P_{\mathrm{b,f}}(z)=\int^{\infty}_{-\infty}|E_{\mathrm{b,f}}(z,t)|^{2}dt$, shown in Fig. \ref{SM1}(b,d), corresponding to Fig. \ref{SM1}(a,c). From these panels, it is seen that $P_{\mathrm{f}}$ transfers to $P_{\mathrm{b}}$ during propagating inside the CBG segment. In panel. \ref{SM1}(b), corresponding to a smaller $r_{1}$, $P_{f}-P_{b}>0$, and becomes constant in UBG segment. According to $V_{\mathrm{pulse}}= v_{g}(1-f)/(1+f)$ ( where $f=P_{\mathrm{b}}/P_{\mathrm{f}}$ \cite{Fu2}), we have $V_{\mathrm{pulse}}>0$. That means the pulse keeps propagating in the forward direction through the setting. While in panel. \ref{SM1}(d), corresponding to a larger $r_{1}$, it is observed that majority of injected power is converted into backward-wave before the pulse reaches the interface. As a result, the pulse bounces back.

To identify the relationship between CBG length and the optical transmission, the transmissivity is investigated for two selected $r_{1}$, with the following expression
\begin{eqnarray}
\eta={\int[|E_{f}(z=L,t)|^{2}+|E_{b}(z=L,t)|^{2}]dt\over\int[|E_{f}(z=0,t)|^{2}+|E_{b}(z=0,t)|^{2}]dt}\times100\%. \label{eta}
\end{eqnarray}
\begin{figure}[tbp]
\centering{\label{fig3a}
\includegraphics[scale=0.38]{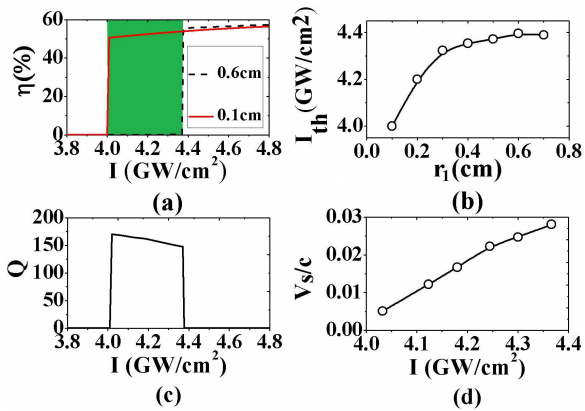}}%
\caption{(Color online) (a) $\eta$ versus $I$, the PI of the input pulse, for different $r_{1}$: 0.1 cm  (red solid curve) and 0.6 cm (black dash curve). (b) $I_{\mathrm{th}}$ vs. $r_{1}$. (c) Factor $Q$ of the AOSD, as a function of incident intensities. (d) Velocity of the BG soliton, namely, $V_{s}$, versus PI, which is roughly estimated by $L/\tau$, where $\tau$ is the time when the soliton propagates from the left end to the right end (i.e. $A\rightarrow B$). Here, in all the panels, we fix $\Delta\delta=7.31$ cm$^{-1}$ and $L=2$ cm.}
\label{SM1_2}
\end{figure}

As an example, Fig. \ref{SM1_2}(a) displays $\eta$, as a function of incident PI for $r_{1}=0.1$ (red solid curve) and 0.6 cm (black dash curve), respectively. From this figure, a threshold value $I_{\mathrm{th}}$ is found for a fixed $r_{1}$ (when $I<I_{\mathrm{th}}$, $\eta\approx0$; while $I>I_{\mathrm{th}}$, $\eta$ rises significatly). In this case, we find: $I_{\mathrm{th}}=$4 GW/cm$^{2}$ for $r_{1}=0.1$cm; $I_{\mathrm{th}}=$4.38 GW/cm$^{2}$ for $r_{1}=0.6$ cm. Note that when $I$ is selected near $I_{\mathrm{th}}$, a nearly stopped soliton propagating in UBG segment can be achieved, see Ref. \cite{Fu1}. Moreover, the relation between $I_{\mathrm{th}}$ and $r_{1}$ is also investigated, see Fig. \ref{SM1_2}(b). As expected, because of the strong pulse retardation in CBG region, the increase of $r_{1}$ also leads to an increase of the $I_{\mathrm{th}}$.

\begin{figure}[tbp]
\centering{\label{fig4a}
\includegraphics[scale=0.3]{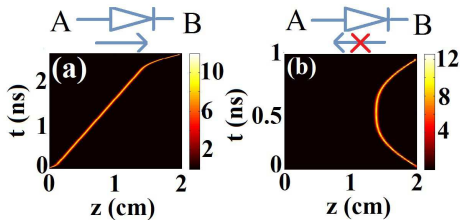}}%
\caption{(Color online)A typical example of an isolator for the setting in Fig. \ref{Model}. The PI of the pulse is $I=4.35$ GW/cm$^{2}$. (a) The pulse injected from side A ($r_{1}=0.1$ cm), which can transmit the setting. For this case, $V_{s}\approx2$ cm/2.43 ns$=0.0274c$. (b) The pulse injected from side B ($r_{2}=0.6$ cm), which is bounced back by the setting. Here, we fix $\Delta\delta=7.31$ cm$^{-1}$ and $L=2$ cm. }
\label{SM2}
\end{figure}

It is interesting to note that, see Fig. \ref{SM1_2}(a), the nonreciprocity can be achieved as long as another CBG segment is added to the right side, namely the swandwich structure in Fig. 1. For example, given $r_{1}=0.1$ cm, $r_{2}=0.6$ cm, an all-optical diode for an unidirectional transmission of soliton could be realized: within 4.0 GW/cm$^{2}$$<I<$4.38 GW/cm$^{2}$, when the pulse is injected into the setting from the left side (namely, side A), it can evolute to a soliton and penetrate the setting with high transmission; while the pulse incident from the right-side end (namely, side B), it is nearly completely blocked.

To identify the quality of the AOSD, we introduce a $Q$ factor of this AOSD as $Q=\eta_{B}/\eta_{A}$\cite{Zhang2014},
where $\eta_{A}$ is the transmissivity of the soliton from $B\rightarrow A$, while $\eta_{B}$ is the transmissivity from $A\rightarrow B$. Fig. \ref{SM1_2}(c) displays $Q$ versus $I$, it shows that a high transmission ratio up to 150 is achieved for the working PI channel (i.e. $I\in[4.0,4.38]$ GW/cm$^{2}$). The real-time evolution of the optical pulse further confirms the high transmission optical diode, see Fig. \ref{SM2}. For the case: $A\rightarrow B$, see Fig. \ref{SM2}(a), not only the pulse maintains its soliton-shape after penetrating right-CBG region, but also majority of pulse power is transmitted, with the transmissivity up to 53\%; while for the case: $B\rightarrow A$, see Fig.\ref{SM2}(b), because of the longer CBG length, the pulse bounced back, with the transmissivity nearly approaching zeros.

Although pulse dispersion is compensated by the induced optical nonlinearity, resulting in an optical soliton, strong retardation of pulse group velocity is still observed, due to the power conversion between forward- and backward-waves in CBG region, see Fig. 2(b). We estimate the soliton propagating velocity from $A\rightarrow B$, with results illustrated in Fig. \ref{SM1_2}(d). It shows that velocity is rather smaller ($<0.03c$) for the working PI channel. Thus the setting indeed can be termed as a slow-light AOSD.

\begin{figure}[tbp]
\centering{\label{fig5a}
\includegraphics[scale=0.3]{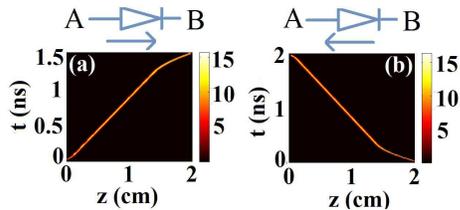}}%
\caption{(Color online)A typical example of a reverse breakdown for the setting in Fig. \ref{Model}. The PI of the pulse is $I=4.73$ GW/cm$^{2}$. (a) The pulse injected from side A ($r_{1}=0.1$ cm), propagating through the total setting. (b) The pulse injected from side B ($r_{2}=0.6$ cm), also penetrating the whole setting. Here, we fix $\Delta\delta=7.31$ cm$^{-1}$ and $L=2$ cm. }
\label{SM3}
\end{figure}

Furthermore, the breakdown effect of an AOSD based on this setting is also investigated. Similar to the breakdown voltage of an electronic diode, when the incident intensity $I>$4.38 GW/cm$^{2}$, the setting becomes two-direction conducting. That means such AOSD is broken dwon, as indicated from the suddenly drop of $Q$, see Fig. 3(c). Fig. \ref{SM3} demonstrates a typical example of such breakdown effect of AOSD. It shows that in both cases (A$\longrightarrow$B; B$\longrightarrow$A), the solitons can penetrate through the setting with certain transmission.

Finally, it should be necessarily pointed out that, in the linear case, pulse would experience serious dispersion and most of the power is reflected since the pulse wavelength is selected near the edge of the bandgap. There is almost no propagating wave exhibited inside the gratings.

Generally, the propagation dynamics of BG solitons can be approximately considered as a quasi-particle system, which cannot exhibit unidirectional transmission in an effective potential \cite{BG7}. However, for the setting demonstrated in this work, a Gaussian optical pulse that is not an exact solution of the BG soliton cannot develop into a BG soliton in CBG segment due to the limited CBG length and pulse intensity, see Fig. \ref{SM1}(b). In this case, the quasi-particle approximation is invalid in this region, and the nonreciprocity of the setting is achieved by introducing two spatially asymmetric CBG segments. We note that, even though the length of CBG is limited, BG solitons can be still generated inside CBG segment by launching Gaussian pulses with high-intensity. In this case, the quasi-particle approximation is re-built in the whole setting, but the diode effect is lost, see Fig. \ref{SM3}.

In conclusion, we have proposed a scheme to realize an AOSD based on a sandwich Bragg-grating structure: built as a left-side chirped grating linked to a uniform one, and again linked to another different chirped grating on the right side. Physical parameters were taken for silicon.  A systematic CMEs has been applied to simulate the evolutions of the pulses propagating dynamics through the setting. The nonreciprocity of the setting is realized by studying the relationship between the optical transmission and CBG length. Based on these analysis, we built an AOSD. A high transmission ratio up to 150 was achieved  when launching a picosecond Gaussian pulse, with the wavelength located near the edge of the photonic bandgap. Because of the balance between the optical nonlinearity and pulse dispersion, such pulses propagate in a form of solitons, which can preserve its soliton-shape during propagating. Moreover, because of the introduction of CBG segment, such solitons propagate at a rather small group velocity. Finally, the breakdown effect of AOSD was also investigated. This setting does not require external assistance such as  radio-frequency modulation, magnetic fields, or optical pumping, which make it possible for the photonic applications, as well as in the all-optical network and devices production, such as optical gate circuit, power isolator, buffer, etc.

This work is supported by the National Natural Science Foundation of China (Grant Nos.11104083, 11204089 and 11205063). Authors appreciate the very useful discussion from Dr. Yikun Liu, Prof. Jianying Zhou and Prof. Boris A. Malomed.

\end{document}